\documentclass[10pt,letterpaper]{article}
\usepackage{opex3}
\usepackage{amsmath}

\begin{document}

\title{A stable, single-photon emitter in a thin organic crystal for application to quantum-photonic devices
}

\author{Claudio Polisseni,  Kyle D. Major, Sebastien Boissier, Samuele Grandi, \\Alex S. Clark$^{*}$, and E. A. Hinds$^{\dag}$}

\address{Centre for Cold Matter, Blackett Laboratory, Imperial College London, London, SW7 2AZ}

\email{$^*$alex.clark@imperial.ac.uk} 
\email{$^\dag$ed.hinds@imperial.ac.uk}



\begin{abstract*}
Single organic molecules offer great promise as bright, reliable sources of identical single photons on demand, capable of integration into solid-state devices. It has been proposed that such molecules in a crystalline organic matrix might be placed close to an optical waveguide for this purpose, but so far there have been no demonstrations of sufficiently thin crystals, with a controlled concentration of suitable dopant molecules.  Here we present a method for growing very thin anthracene crystals from super-saturated vapour, which produces crystals of extreme flatness and controlled thickness. We show how this crystal can be doped with a widely adjustable concentration of dibenzoterrylene (DBT) molecules and we examine the optical properties of these molecules to demonstrate their suitability as quantum emitters in nanophotonic devices.  Our measurements show that the molecules are available in the crystal as single quantum emitters, with a well-defined polarisation relative to the crystal axes, making them amenable to alignment with optical nanostructures. We find that the radiative lifetime and saturation intensity vary little within the crystal and are not in any way compromised by the unusual matrix environment. We show that a large fraction of these emitters are able to deliver more than $10^{12}$ photons without photo-bleaching, making them suitable for real applications.\\ 
\\
\end{abstract*}

\bibliography{OpEx_Claudio_OvercoatarXiv}
\bibliographystyle{osajnl}

\section{Introduction}

Single photons are important as a means of transporting quantum information because they interact weakly with their environment and can carry information in many degrees of freedom \cite{OBrien:07}. In order to scale up the complexity of quantum operations, there is now a real need for a source of identical photons, promptly delivered on demand, and ideally integrated in the solid state into a photonic chip. Quantum dots are one promising approach, but despite significant recent progress \cite{HePRL:13}, photons produced by different dots are generally distinguishable. Nitrogen-vacancy (N-V) centres in diamond are also promising \cite{Kurtsiefer:00}, though it is challenging to place colour centres at specific sites in a photonic circuit \cite{Aharonovich:11} and the emitted photons have uncertain frequencies due to phonon-assisted sidebands and the low probability of emission on the zero-phonon line. By contrast, dibenzoterrylene (DBT) molecules embedded in anthracene at cryogenic temperatures produce a high yield of indistinguishable photons \cite{Nicolet1:07} in a 30~MHz-wide line at 785~nm \cite{Tamarat:00,Trebbia:09}.  It has been proposed that these molecules could deliver photons very efficiently into a nearby waveguide \cite{Hwang:11, Verhart:14}.

\begin{figure}[t!]
\centering\includegraphics[width=12cm]{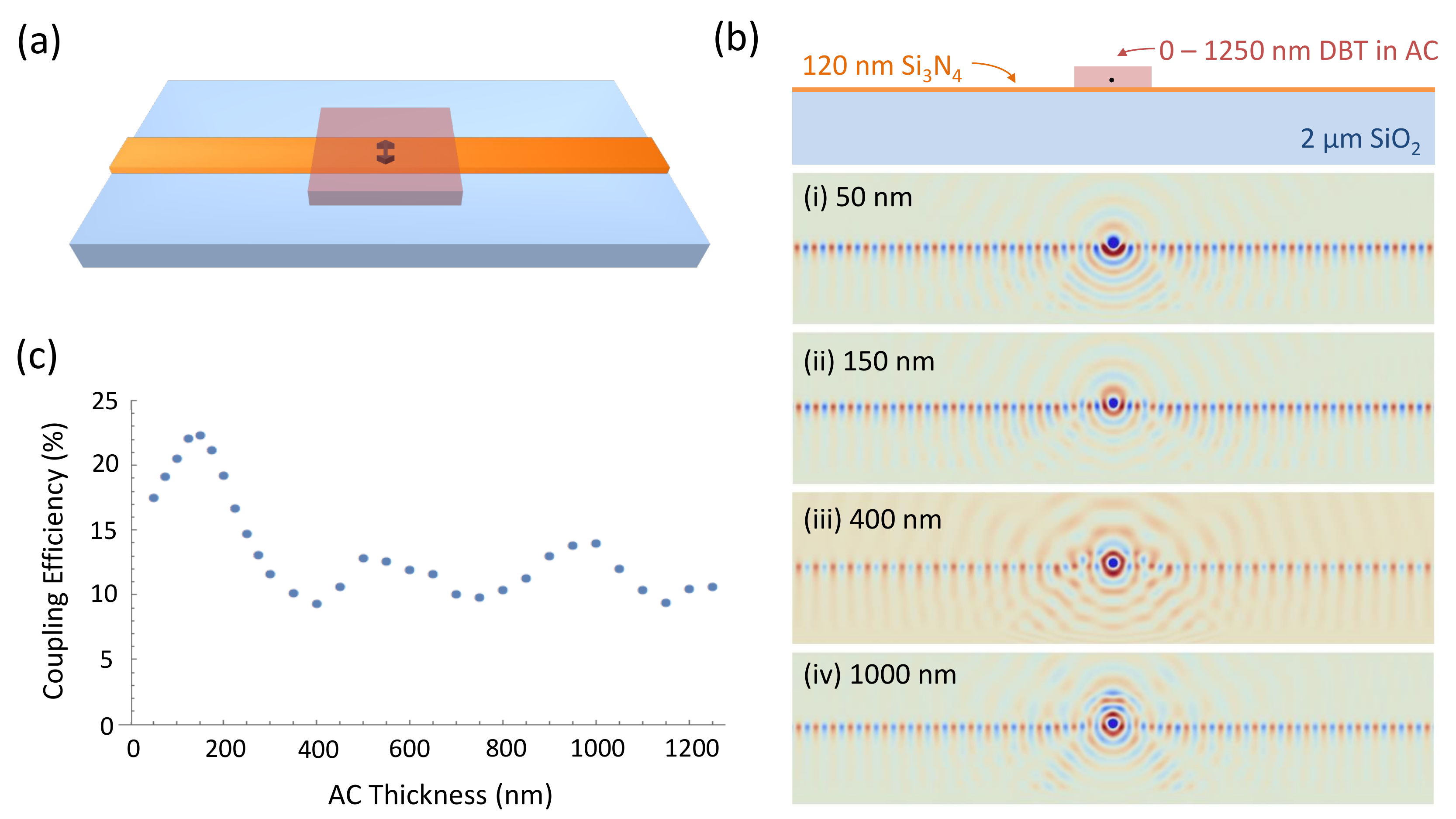}
\vspace{0.5cm}
\caption{We have numerically solved Maxwell's equations for a radiating dipole embedded in a $2\,\mu$m-square ``crystal" of anthracene. The dipole lies 32\,nm above a silicon nitride waveguide  500\,nm wide and 120\,nm thick on a glass substrate. (a) Sketch of the geometry. (b) Top: side view of substrate, waveguide, anthracene and dipole. Below: Simulation results for several thicknesses of the crystal. (i) Radiation is lost into the substrate. (ii) Radiation propagates in the waveguide. (iii-iv) Radiation is lost into the thick anthracene layer. (c) Graph showing the calculated coupling efficiency as the thickness of the anthracene crystal is varied. This peaks at $\sim 22\%$ (total in both directions) for a thickness of $\sim 150$ nm.
}
\label{Fig:FDTD}
\end{figure}

In pursuit of that idea, we have used the open-source MEEP package \cite{Oskooi:10} to solve Maxwell's equations for a radiating dipole located 32\,nm above a silicon nitride waveguide and embedded in a $2\,\mu$m-square ``crystal" of anthracene, as illustrated in Fig.~\ref{Fig:FDTD}(a). The top panel in Fig.~\ref{Fig:FDTD}(b) illustrates this from the side, and plots the field radiated by the molecule for several different thicknesses of crystal. Unless the crystal is thin, much of the radiation goes into the anthracene instead of the waveguide because the anthracene refractive index -- $n\simeq 1.8$ for the polarisation of interest -- is almost as high as the $n=2.0$ of the silicon nitride. Fig.~\ref{Fig:FDTD}(c) plots the efficiency of coupling into the waveguide as a function of the anthracene thickness and shows an optimum thickness of $\sim150\,$nm. To make such a crystal, we could not use the normal co-sublimation methods \cite{Nicolet1:07,Trebbia:09,Nicolet2:07,Makarewicz:12,Major:15}, which yield crystals that are $1-2\mu\textrm{m}$ thick.  Films of the desired thickness were made by spin coating a solution of DBT and anthracene, and these are known to give stable photon emission from the DBT \cite{Toninelli:10}, but we found that these films are always too rough for our purpose, varying in thickness by as much as 50\% over regions only a few microns across.

In this paper we show how to grow a uniformly thin crystal of anthracene containing a controlled density of DBT dopant molecules, and we demonstrate that these molecules are suitable for use as single photon sources on a chip. Section~\ref{sec:growth}  presents our methods for growing and doping the crystal. Section~\ref{sec:microscopy} describes the microscopy used to examine the crystals, image the molecules and investigate the molecular polarisation. In Sec.~\ref{sec:DBToptics} we characterise the antibunching and saturation of the fluorescence and we explore the photobleaching of the molecules. Section~\ref{sec:conclusions} provides a concluding summary of our results.

\section{Growing the crystals}
\label{sec:growth}

\begin{figure}
\centering\includegraphics[width=12cm]{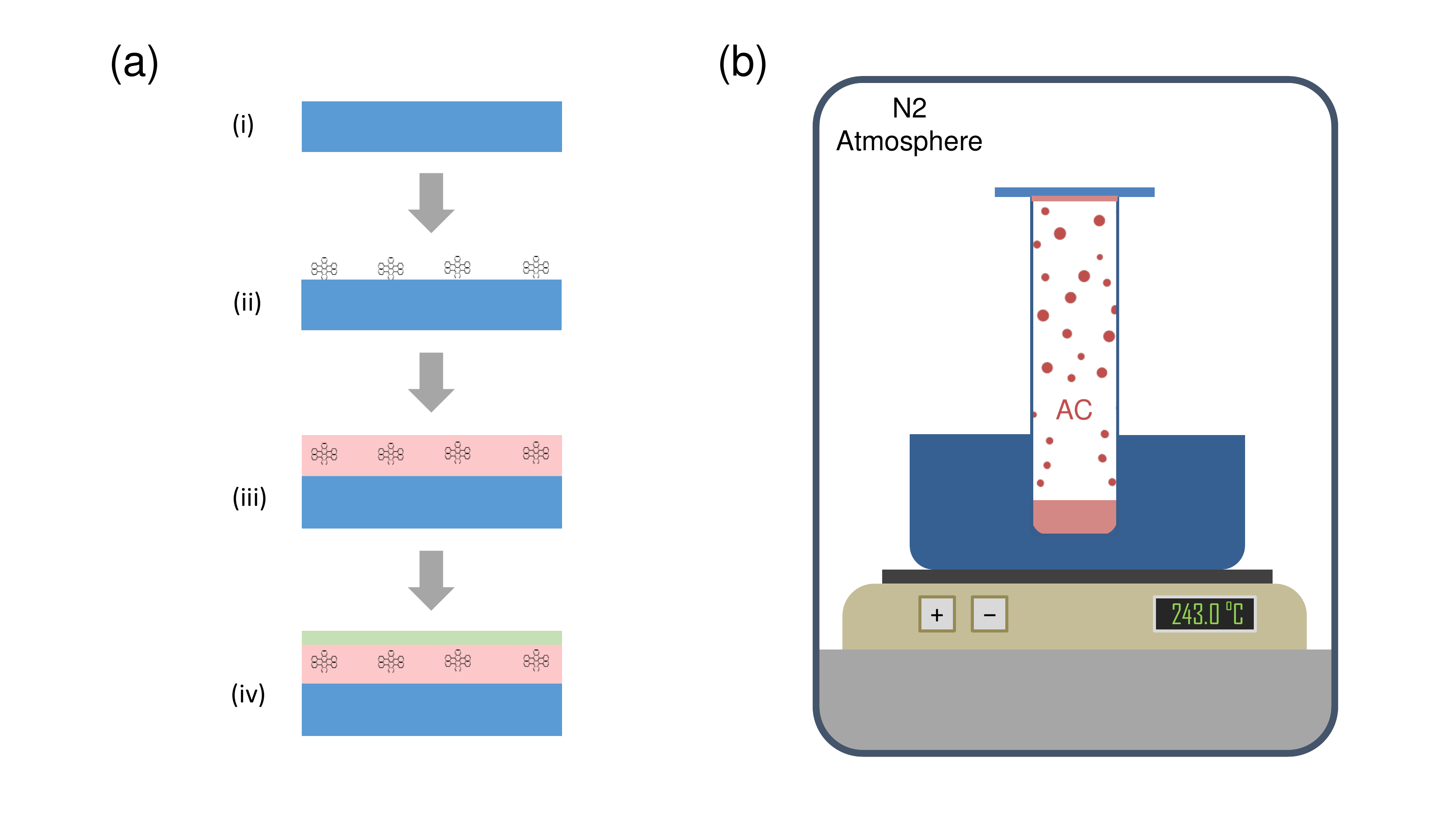}
\vspace{-0.5cm}
\caption{Preparing the thin crystal of anthracene doped with DBT.  (a) Steps involved in the processing. (i) clean of the glass substrate in a plasma, (ii) spin-coat DBT molecules onto the surface, (iii)  grow a thin anthracene crystal over the molecules, (vi) spin-coat a protective layer of PVA. (b) The crystal growth apparatus is a test tube with anthracene powder heated at the bottom and the glass substrate covering the open top, all in a nitrogen atmosphere inside a glove bag.}
\label{Fig:growth_process}
\end{figure}

Figure \ref{Fig:growth_process}(a) illustrates the steps we follow to produce suitable crystals. We start with (i), a glass cover slip that has been cleaned in an oxygen plasma. Next, we make a 1 mMol solution of the DBT molecules\footnote{7.8,15.16-dibenzoterrylene supplied by the Institute for PAH-Research, Greifenberg, Bavaria} in toluene, which we further dilute in diethyl ether at a ratio of 1:1000 (volume) to reach the desired concentration. In step (ii) we spin-coat the cover slip with  $20\,\mu\mathit{l}$ of this solution, ramping up to 2000 rpm over 5 s, then holding for 60 s before ramping down suddenly. This recipe gives a surface coverage of approximately 0.4 DBT molecules per square micron. In step (iii) we grow a thin anthracene crystal on the cover slip,  inside a glove bag purged of air and filled with nitrogen, as sketched in Fig. \ref{Fig:growth_process}(b).  A temperature-controlled copper block heats the bottom end of a glass vial, containing $3\,$g of anthracene powder, to a temperature $T_b$ that we can set. Once the bottom temperature has stabilised, the DBT-coated cover slip is placed on top of the tube for approximately 1 minute to grow the crystal. During this growth period, the temperature at the top is $T_t\simeq25\,^{\circ}$C. Finally, in step (iv), we mix 4\% by weight of PVA powder in deionised water at $\sim65\,^{\circ}$C and wait (about an hour) until the powder is dissolved. We spin $20\,\mu\mathit{l}$ of this solution on top of the anthracene, ramping up to 3000 rpm and back down again over a minute. This covering is necessary because uncoated anthracene sublimes and is gone in less than a day, whereas the coated crystals last for months (at least) at room temperature.

\begin{figure}[t]
\centering\includegraphics[width=13cm]{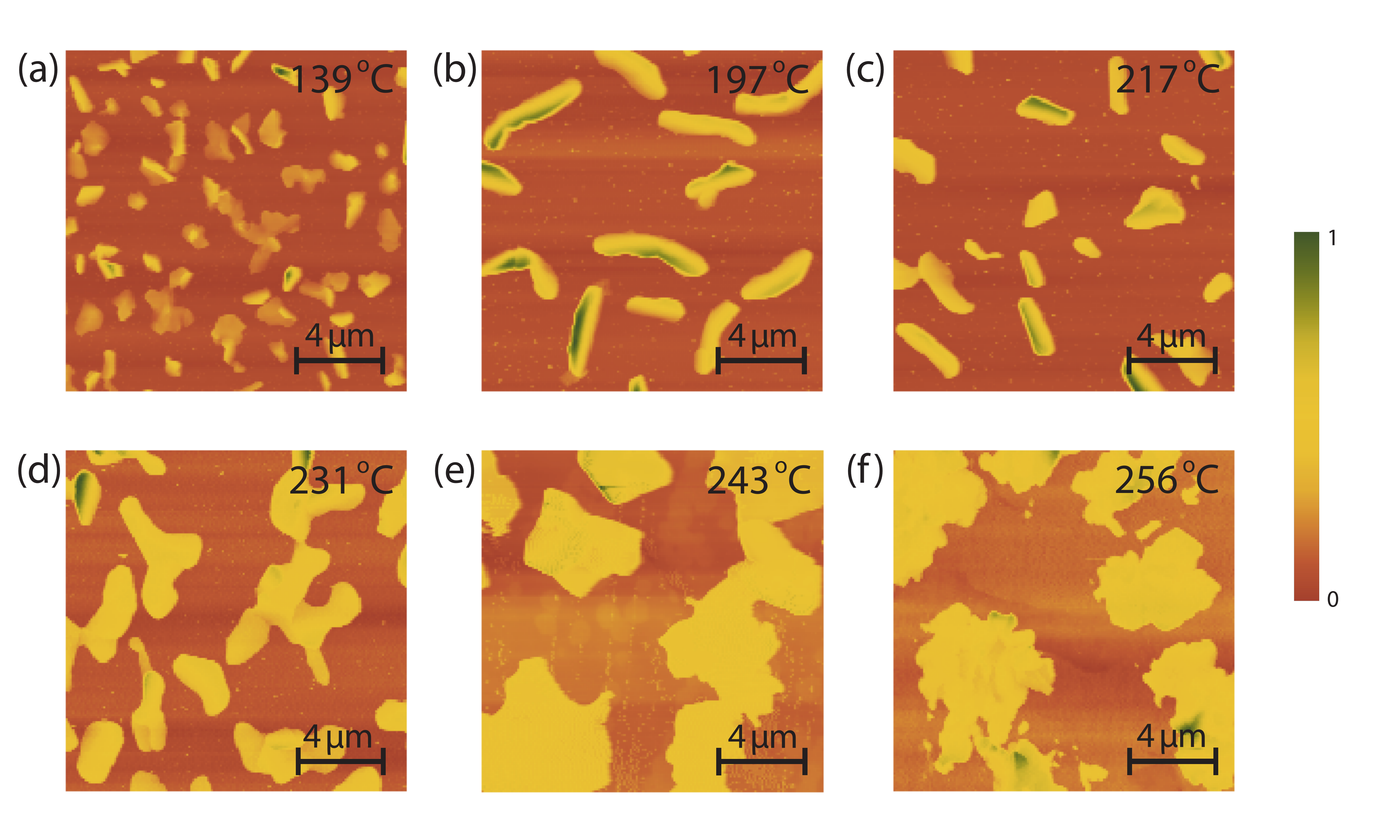}
\caption{Atomic force microscope images  (taken in tapping mode) of the crystals grown at a variety of bottom temperatures $T_b$ (a) $139\,^{\circ}$C, (b) $197\,^{\circ}$C, (c) $217\,^{\circ}$C, (d) $231\,^{\circ}$C, (e) $243\,^{\circ}$C, and (f) $256\,^{\circ}$C. These images show the transition from tall, thin crystals - similar to those reported in \cite{Major:15} - to wide, flat mesas. The transition occurs between $T_b=220$ and $240\,^{\circ}$C.}
\label{Fig:AllPics}
\end{figure}

\begin{figure}[b]
\hspace{-1.5em}
\includegraphics[width=14cm]{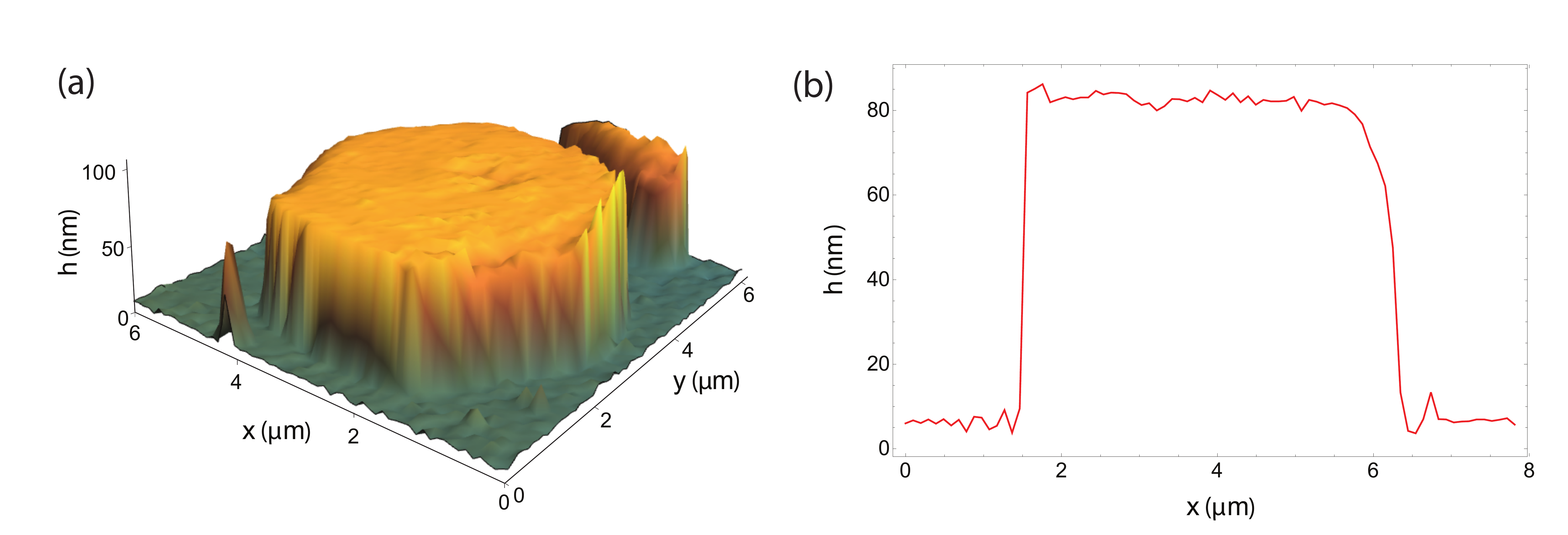}
\caption{Atomic force microscope image of a thin anthracene mesa. (a) 3D plot. (b) Section through the centre along $x$, showing  a flat-top profile with a height of $\sim$80~nm and very low roughness.}
\label{Fig:AFM}
\end{figure}

 Changing $T_b$ has a profound effect on the morphology of the crystals that grow, as we demonstrate in  Fig.~\ref{Fig:AllPics} -- a set of atomic force microscope images showing crystals grown at successively higher values of the bottom temperature. With $T_b\le218\,^{\circ}$C (Figs.~\ref{Fig:AllPics}(a-c)), the crystals are long and thin in cross section and stick up from the surface as 3-dimensional structures, reminiscent of those described in \cite{Major:15}. By contrast, when $T_b\ge243\,^{\circ}$C (Figs.~\ref{Fig:AllPics}(e, f)), they change to wide, flat mesas, whose height grows roughly linearly with the deposition time. A typical mesa is shown in Fig.~\ref{Fig:AFM}(a), together with a cut through the centre along $x$, plotted in Fig.~\ref{Fig:AFM}(b). This particular crystal was grown for $50\,$s at $T_b=243\,^{\circ}$C, and is $80\,$nm thick. The roughness measured on the top of the mesa is  $1.1\,$nm rms, but that is about the level of the instrumental noise, so we believe the crystal is smoother than $1\,$nm. Thus, by controlling the temperature $T_b$ and the growth time, we have produced crystals of very uniform thickness in the range $40 - 150\,$nm, a suitable range for coupling the DBT molecules to a waveguide.

Before turning to the optical properties of the DBT molecules underneath this thin crystal, we digress briefly to comment on the thermodynamics of the crystal growth. The pressure $p_{t}$ of anthracene vapour at the top of the test tube, from which the crystal grows, is controlled by the temperature at the bottom, where we can reasonably suppose  the vapour is in equilibrium with the solid, at the vapour pressure $p_0(T_{b})$. Thus, taking into account the temperature gradient along the test tube, and treating the anthracene vapour as an ideal gas, we expect $p_t=\sqrt{T_t/T_b}p_0(T_{b})$. The chemical potential of this (ideal) vapour at the top of the test tube is $\mu_1=-k_B T_t\ln[p_Q/p_t]$ (with quantum pressure $p_Q=(\frac{m k_B T_t}{2\pi \hbar^2})^{3/2}k_B T_t$). Let us compare this with the chemical potential of a fully covered anthracene surface at temperature $T_t$, for which the vapour pressure is $p_0(T_{t})$, and therefore the chemical potential is $\mu_2=-k_B T\ln[p_Q/p_0(T_{t})]$.  The chemical potential difference is $\mu_1-\mu_2=k_B T_t\ln[p_t/p_0(T_t)]$. We see that when the pressure at the top of the test tube is equal to the vapour pressure for anthracene at temperature $T_t$, there is no chemical potential difference. Then the crystal sits in equilibrium with the vapour and cannot grow, which is the case when $T_b=T_t$.

\begin{figure}[b]
\centering\includegraphics[width=13cm]{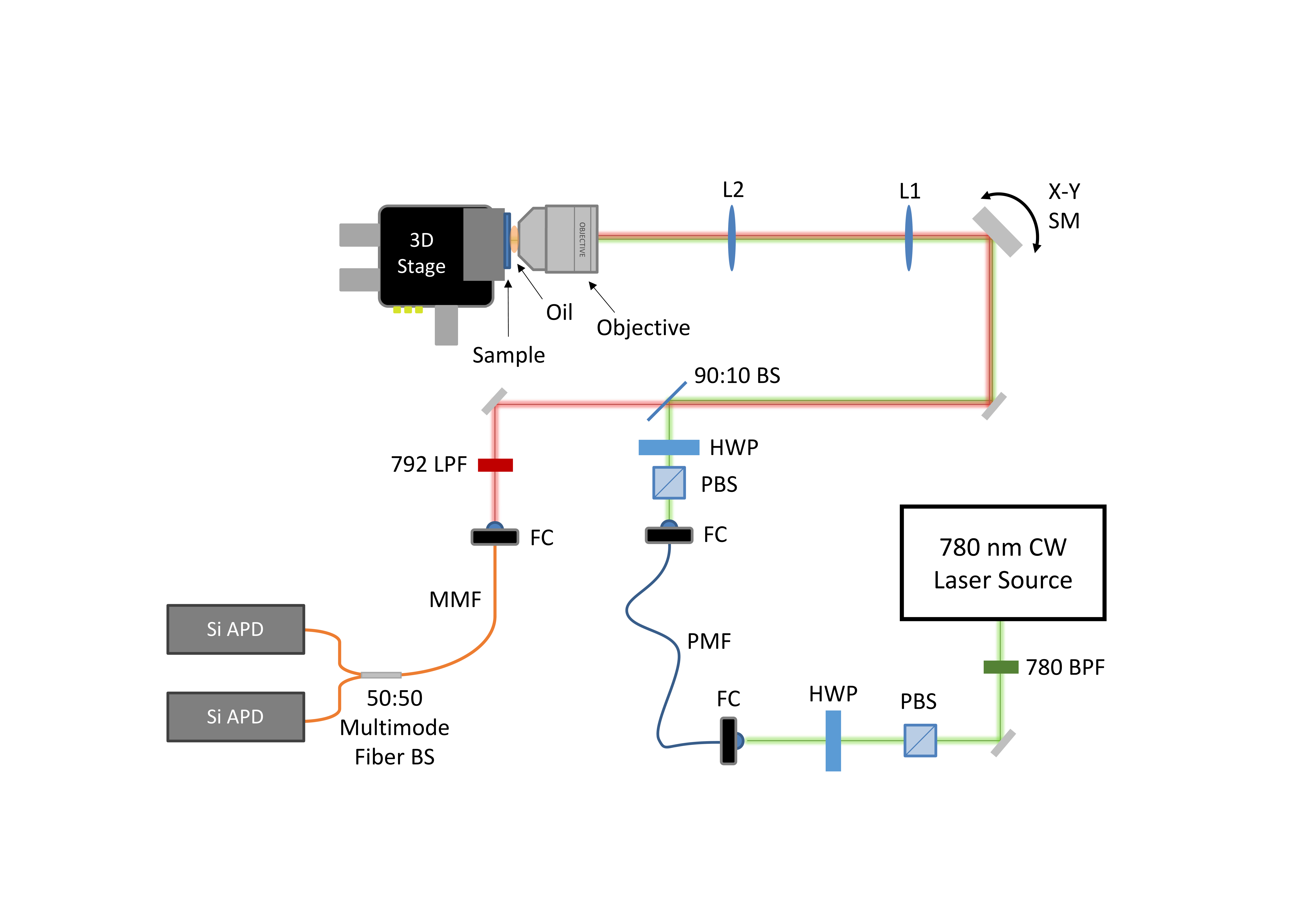}
\caption{Schematic of our confocal microscope system. 780-BPF: 780~nm band pass filter; PBS: polarising beam splitter; HWP: half wave plate; FC: fiber coupler; PMF: polarisation maintaining fiber; 90:10 BS: 90\% transmission, 10\% reflection cube beam splitter; X-Y SM: X-Y steering mirrors; L1, L2: telecentric lens system $f_1=75\,$mm, $f_2=250\,$mm; Objective: Nikon OFN25 DIC N2 $60\times$ microscope objective; 3D Stage: Thorlabs Nanomax MAX311/M translation stage for positioning the sample ; 792-LPF: Chroma technology RET792LP long pass filter; MMF: multimode fiber; 50:50 Multimode Fiber BS: 50\% transmission, 50\% reflection multimode fiber beam splitter; Si APD: Perkin Elmer SPCM-AQRH-15-FC silicon avalanche photodiode detectors.}
\label{Fig:oil_confocal}
\end{figure}

A rise of $T_b$ increases the vapour pressure at the bottom, which drives up the pressure at the top, and hence increases the chemical potential difference between the crystal and the vapour at the top. When this potential difference is small, the crystal that grows is constrained in its morphology by the need to minimise its surface energy (see \cite{Grimbergen:98,Verlaak:03}, for example) but as more free energy becomes available other conformations become possible. Above a critical value, the crystal is able to grow in a two-dimensional way \cite{Verlaak:03}, and this is the origin of the transition that takes place somewhere between $220$ and $240\,^{\circ}$C. At these temperatures, the ratio $p_0(T_b)/p_0(T_t)$ of vapour pressures is in the range  $(0.4- 3.4)\times 10^7$ \cite{Chen:06,Lide:09,WebBook}, and the corresponding chemical potential difference between the vapour at the top and the infinite anthracene crystal is $0.41(2)\,$eV. Following section 4 of \cite{Sassella:06}, we expect this transition to occur at a chemical potential difference of $2a b(2\gamma-\sigma)$, where $a b$ is the area of $(001)$ plane of the unit cell, while $\gamma$ and $\sigma$ are the surface energies per unit area of the (001) plane and the interface with the substrate, respectively. From \cite{Northrup:02} the values are $a b=51.7\,\AA^2$, $\gamma=3.3\,$meV/$\AA^2$. Comparing this theory with our measurement, we conclude that the energy per unit area of the anthracene binding to the glass is $\sigma=2.6(2)\,$meV/$\AA^2$. We are not aware of any previous determination of this quantity, but note that it is of similar magnitude to the value taken by \cite{Verlaak:03} for other -acenes.

\section{Microscopy}
\label{sec:microscopy}

\begin{figure}[b]
\centering\includegraphics[width=9cm]{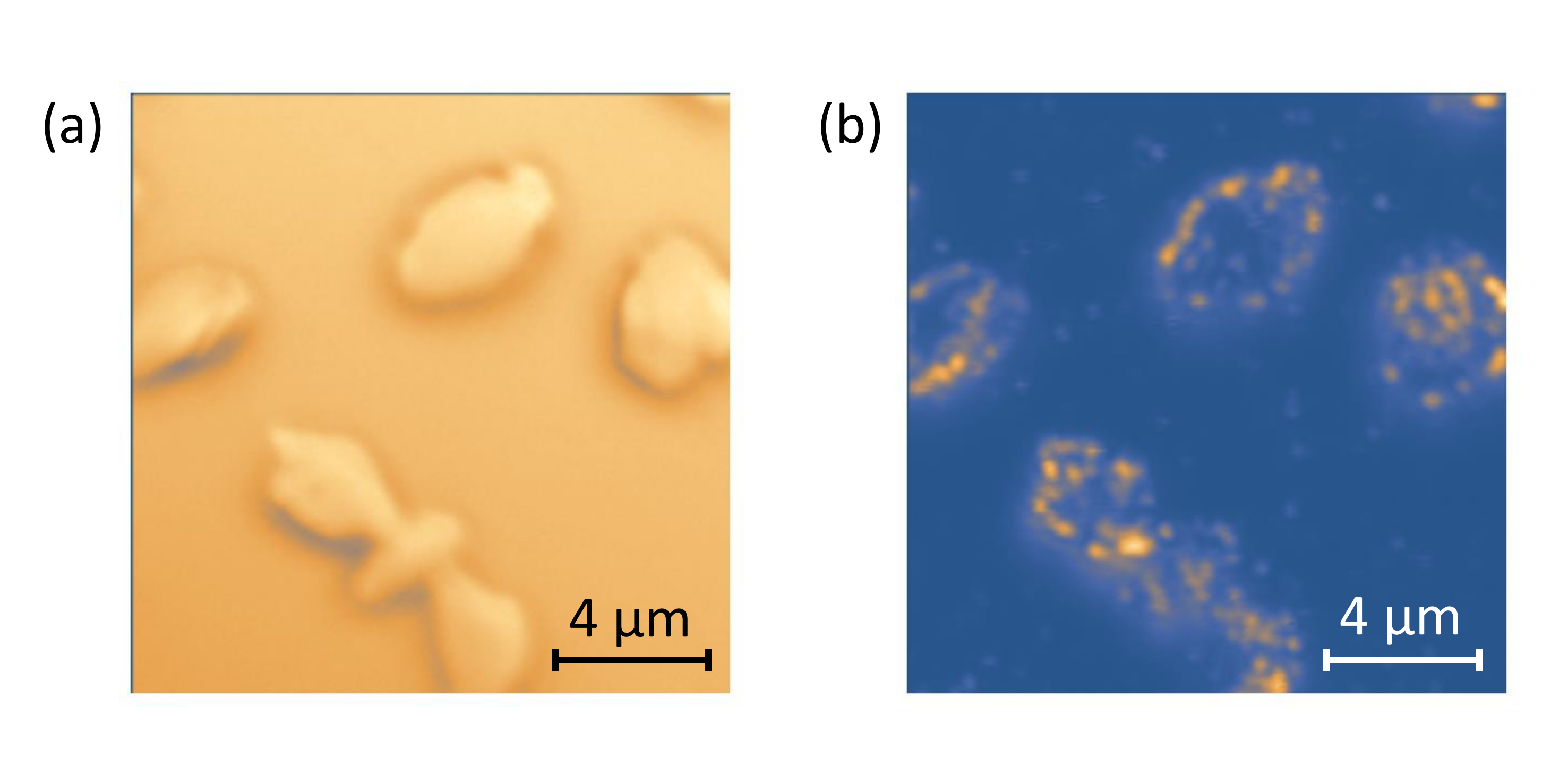}
\caption{Confocal microscope images. (a) Anthracene crystals viewed in reflected excitation light (long-pass filter removed). (b) The same crystals viewed with the long-pass filter in place. Nearly all the spots are single DBT molecules}
\label{Fig:confocal}
\end{figure}

We turn now to the analysis of the DBT molecules embedded under these thin anthracene crystals, which we image using the confocal microscope setup shown in Fig.~\ref{Fig:oil_confocal}. The light source is an external-cavity continuous-wave diode laser (TEC 100 Littrow - Lynx, Sacher Lasertechnik) producing $8\,$mW at $780\,$nm, filtered by a $10\,$nm-wide bandpass (BPF) to suppress any amplified spontaneous emission at wavelengths longer than $785\,$nm. The light is linearly polarised (PBS) and coupled into a polarisation-maintaining, single-mode fibre (PMF), which filters the spatial mode.  A half-wave plate (HWP) allows us to adjust the polarisation axis so that the light couples only to one of the principle axes of the fibre. The output from the fibre is polarised once again (PBS) and the axis is then adjusted by a second half-wave plate. A polarisation-insensitive beam splitter directs 10\% of the light ($\sim0.5\,$mW) onto two steering mirrors (SM) that raster-scan the beam direction. Lenses L1 and L2  expand the beam size from $2\,$mm to $6.7\,$mm and ensure that it always enters the microscope objective centrally while the angle is being scanned. We use an infinity-corrected, Plan-Apochromat, oil-immersion objective lens. This has  $60\times$ magnification with a numerical aperture of 1.40, and is corrected for 170~$\mu$m cover slips with a working distance of 130~$\mu$m. On fitting the focussed spot to a Gaussian, we typically measure a full width at half maximum intensity of $400\,$nm in the object plane. The steering mirrors allow us to scan this spot over an area of $20\times20\,\mu$m.

The $780\,$nm laser excites molecules from the ground state $S_{0}$ to the first excited singlet $S_{1}$. Although the laser frequency lies  $2\,$THz above the $S_{0,0}-S_{1,0}$ ``zero-phonon" resonance of DBT, the homogeneous linewidth at room temperature is approximately $10\,$THz, so molecules are excited both to the $S_{1,v=0}$ state at the bottom of the upper band and to vibrationally excited states $S_{1,v}$ of the upper electronic level. These relax very rapidly to $S_{1,0}$, which then decays radiatively back to the ground level $S_{0}$. The ``zero-phonon" light is emitted at a wavelength of $\sim785\,$nm, while the transitions to vibrationally excited ground states $S_{0,v}$ produce sidebands further to the red. This fluorescence is collected by the microscope and retraces the path of the incident light back as far as the beam splitter, where 90\% of the photons are transmitted. A $792\,$nm long-pass filter (LPF) rejects scattered excitation light -- and also rejects DBT fluorescence close to the zero-phonon line -- but transmits the more red-shifted sidebands of the fluorescence. This light is coupled into a multimode fibre (MMF) and a fibre beam splitter delivers half the power to each of two silicon avalanche photodiode detectors (Si-APD).

With the long-pass filter removed, there is a strong signal in the detectors from back-scattered $780\,$nm excitation light. Figure~\ref{Fig:confocal}(a) shows six anthracene crystals, viewed in this way after attenuating the excitation power to $\sim1\,$nW. The image in Fig.~\ref{Fig:confocal}(b), taken with the long-pass filter in place and with the attenuator removed,  shows the individual DBT molecules embedded in the same crystals. These scans cover an area of $16\times16\,\mu$m and have a resolution of  $200\times200$ pixels.

\begin{figure}[b]
\centering\includegraphics[width=13cm]{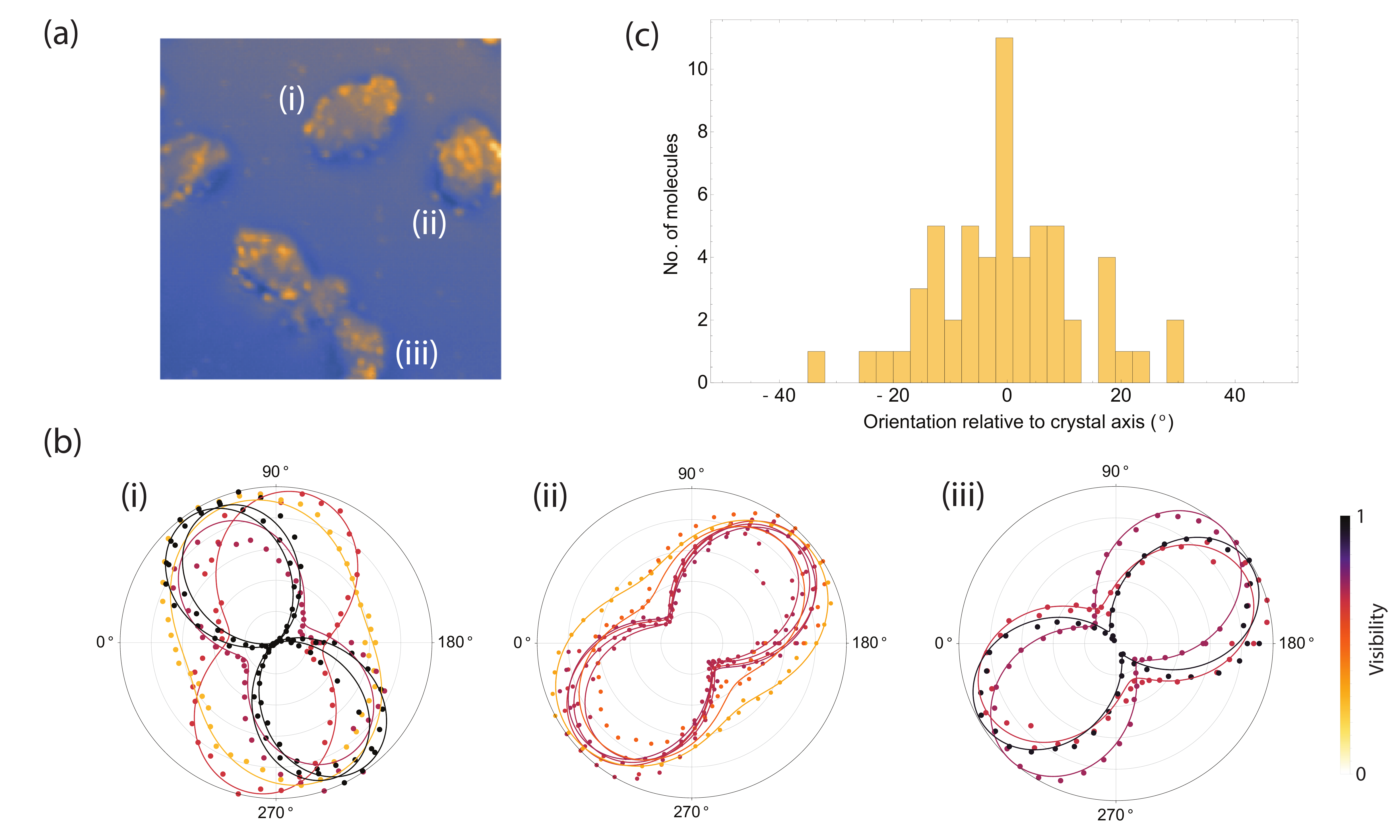}
\caption{Test of DBT polarisation. (a) Sum of the two images from Fig.~\ref{Fig:confocal}. (b) Normalised polar plots of molecule emission as the excitation laser polarisation angle is varied. Labels (i-iii) indicate the crystal in which the molecules reside, as shown in (a). The colour of the lines signifies the visibility of the fitted function to the data. Within a given crystal the optical dipoles clearly share a common orientation. (c) Histogram showing the spread of molecule orientations relative to the mean orientation, taken over 58 molecules in 12 crystals.}
\label{Fig:polarization_confocal}
\end{figure}

In order to have good control over the density of photo-emitters, we first deposit the DBT molecules on the glass surface then over-coat them with anthracene, as described in section 2. We therefore needed to establish whether the molecules remain stuck on the surface, underneath the crystal, or are incorporated into the  anthracene matrix. DBT molecules occupying the main site in anthracene have the long axis of the terrylene moiety - and hence the $S_0 - S_1$ optical dipole moment - polarised along the b-axis of the crystal \cite{Nicolet2:07}, whereas they must have a random orientation when we first deposit them on the glass. Thus, our first test was to measure the fluorescence as a function of the polarisation angle of the excitation light, to see whether the DBT molecules in a given crystal share a common orientation. Figure~\ref{Fig:polarization_confocal}(a) sums the two images in Fig.~\ref{Fig:confocal}, to show the DBT molecules and the anthracene crystals simultaneously, and indicates three crystals, labeled (i - iii), where we have measured the DBT polarisation in some detail. In each crystal we identify molecules that are well isolated from each other, and for those molecules we measure the fluorescence intensity, normalised to incident power, as a function of the polarisation angle $\theta$. We determine the background level by measuring the signal from a nearby place on the same crystal where there are no DBT molecules, and this is subtracted. The polar plots in Fig.~\ref{Fig:polarization_confocal}(b) show the angular dependences measured in crystals (i - iii).  In crystal (i) four out of the five molecules exhibit the same orientation, though one of these has quite poor contrast. In crystal (ii) all six molecules have the same orientation, as do the three measured in crystal (iii).  The lines drawn in Fig.~\ref{Fig:polarization_confocal}(b) are fits of the form $A\cos^{2}(\theta+\phi)+B/(A+B)$, with $\phi$ being the direction of the molecular alignment. In all, we looked at 58 molecules in 12 crystals and found, as summarised in Fig.~\ref{Fig:polarization_confocal}(c), that the DBT molecules in a given crystal are indeed closely aligned with each other. This provides strong evidence that the molecules move inside the anthracene as it grows and become aligned with the b-axis of the crystal. It is possible that some of the dispersion in orientation is associated with DBT molecules that occupy the secondary insertion site, known as the red site, where the orientation is rotated by $7^{\circ}$ \cite{Nicolet2:07}, but we have not yet explored that possibility.

\section{Optical properties of the DBT}
\label{sec:DBToptics}

In order to prove that the DBT fluorescence seen in Fig.~\ref{Fig:confocal}(b) comes from individual molecules, we measure the correlation between photon arrival times in the two photodetectors (Fig.~(\ref{Fig:oil_confocal})) using a time-correlated counting card (TCCC, Picoquant Timeharp200) and a 755~nm, 83~ps pulsed laser (Picoquant, PDL-800B). This determines the average probability of detecting a photon in the second detector during a narrow time window $\delta t$ centred on time  $t+\tau$, having previously registered one in the first detector at time $t$. For small enough $\delta t$ this is the discrete approximation to the (unnormalised) second-order correlation function $C^{(2)}(\tau)$ \cite{MigdallBook:13}

\begin{equation}
C^{(2)}(\tau) =\langle \hat{a}^\dagger(t)\hat{a}^\dagger(t+\tau)\hat{a}(t)\hat{a}(t+\tau) \rangle ,
\end{equation}

\noindent where $\hat{a}^\dagger$ and $\hat{a}$ are the photon creation and annihilation operators, respectively.
With $\delta t=106.9\,$ps and an excitation power of 16~$\mu$W we have count rates of order $4\times 10^4\,\mbox{s}^{-1}$ in each detector. Because the pulse repetition time $\Delta t=25\,$ns is substantially longer than the decay time $T_1$ of the excited state, we expect the correlation we measure to be well described by the function

\begin{equation}
C^{(2)}(\tau) \simeq B+ N\sum_{n=-\infty}^{\infty}(1-\frac{\delta_{0n}}{m})  e^{-\frac{|\,\tau-n\,\Delta t |}{T_1}}\,,
\label{eq:g2Function}
\end{equation}
where $B$ is a background due to dark counts in the detector, $N$ is a normalising factor, $n$ is an integer that labels the excitation pulses,  $\delta_{0n}$ is the Kronecker delta, and $m$ is the number of molecules emitting the light.

\begin{figure}[t]
\centering\includegraphics[width=9cm]{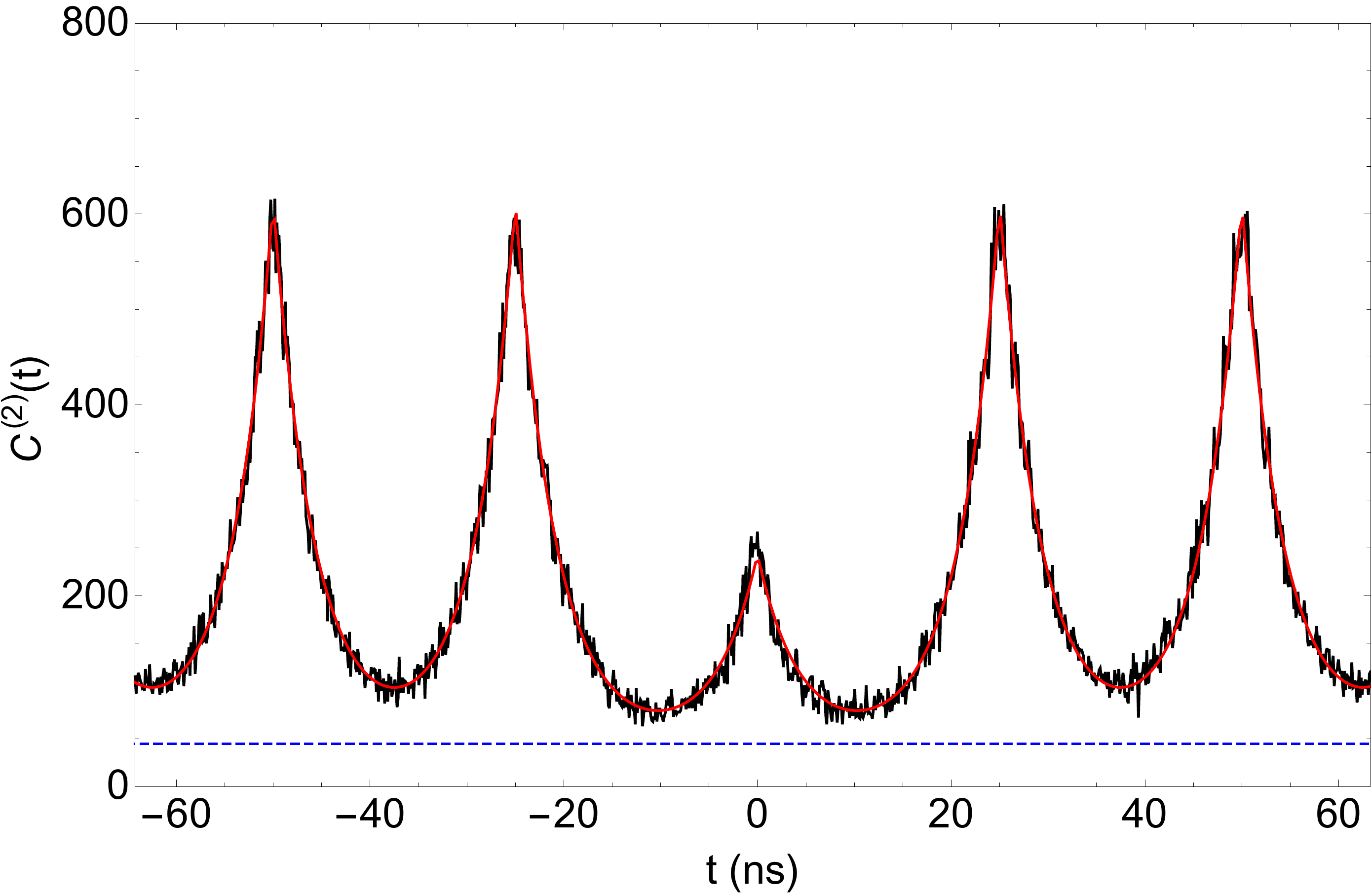}
\caption{Second order correlation function of the light from a single fluorescent spot within a thin anthracene crystal when pumped with a pulsed laser.  The reduced peak in the centre shows that this is a single DBT molecule. Black line: data accumulated over 30 minutes with the timing resolution set to $106.9\,$ps. Red line: fit to Eq.(\ref{eq:g2Function}). Blue dashed line: background level deduced from the fitted function.}
\label{Fig:g2}
\end{figure}

The black line plotted in Fig.~\ref{Fig:g2} shows the experimental result obtained after 30 minutes of measurement, while the red line is the fit of Eq.~(\ref{eq:g2Function}) to this measurement. We see that the central peak is strongly suppressed because we are observing a single molecule that is unable to scatter two photons at once. The fit gives the number of molecule as $m=1.53(15)$, or in terms of the normalised second-order correlation function, $g^{(2)}(0)=1-1/m=0.346(59)$. We interpret this to mean that we are observing a single molecule, but that we also collect a small amount of light from other nearby molecule(s) in the periphery of the field of view. By repeating this experiment on many different fluorescent spots we have established that virtually all the emitters are single molecules when the DBT density is low, as in Fig.~\ref{Fig:confocal}. At higher density, we also see bright spots with higher central $C^{(2)}$ peaks, corresponding to clusters of two or more DBT molecules. Returning to Fig.~\ref{Fig:g2}, the blue dashed line shows the background level $B=44.7(2)$.  The same fit  gives the excited state lifetime as $T_1=4.23(5)\,$ns, and this is typical of the lifetimes for all the emitters. It is also similar to the lifetimes measured in thick crystals \cite{Nicolet1:07,Tamarat:00,Trebbia:09,Major:15} and in spin-coated thin films \cite{Toninelli:10}.

\begin{figure}[b]
\centering\includegraphics[width=9cm]{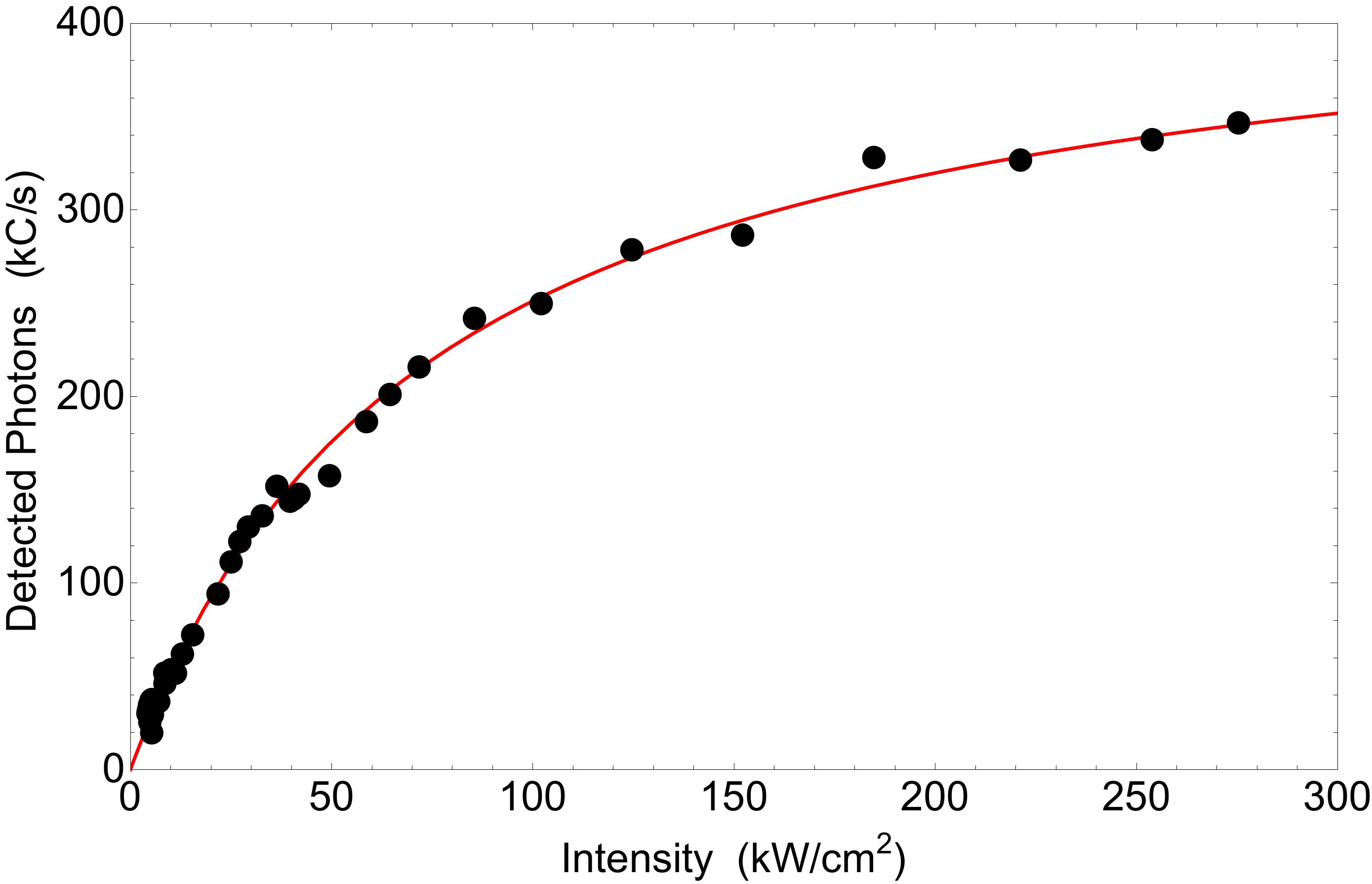}
\caption{A typical plot of detected molecule emission for various illumination intensities, showing a saturation intensity of $I_{sat}=75(3)~$kW/cm$^{2}$ and a maximum count rate of $R_{max}=440~$kC/s. Red line: fit to the data using Eq.~\ref{Eq:sat}.}
\label{Fig:saturation}
\end{figure}

When this same molecule is driven continuously by a laser on resonance, the scattering rate in steady state is given by the optical Bloch equations  \cite{LoudonBook:00} as
\begin{equation}
R = \frac{I}{I+I_{sat}} R_{max}\,,
\label{Eq:sat}
\end{equation}
where $I$ is the laser intensity at the molecule, $I_{sat}$ is the saturation intensity, and $R_{max}$ is the limiting rate at high intensity. The saturation intensity is linked to the Rabi frequency $\Omega$ and to the relaxation times by $I/I_{sat}=\Omega^2 T_1 T_2$. We have determined the saturation intensity of this molecule by measuring the scattered photon rate as a function of the incident intensity, as shown by the points plotted in Fig.~\ref{Fig:saturation}. The red line is a fit of Eq.~(\ref{Eq:sat}) to the data, which gives the result $I_{sat}=75(3)\,$kW/cm$^2$. We are not aware of any other determination of this parameter, but note that closely related data are also presented in \cite{Toninelli:10}.

\begin{figure}[b]
\centering\includegraphics[width=9cm]{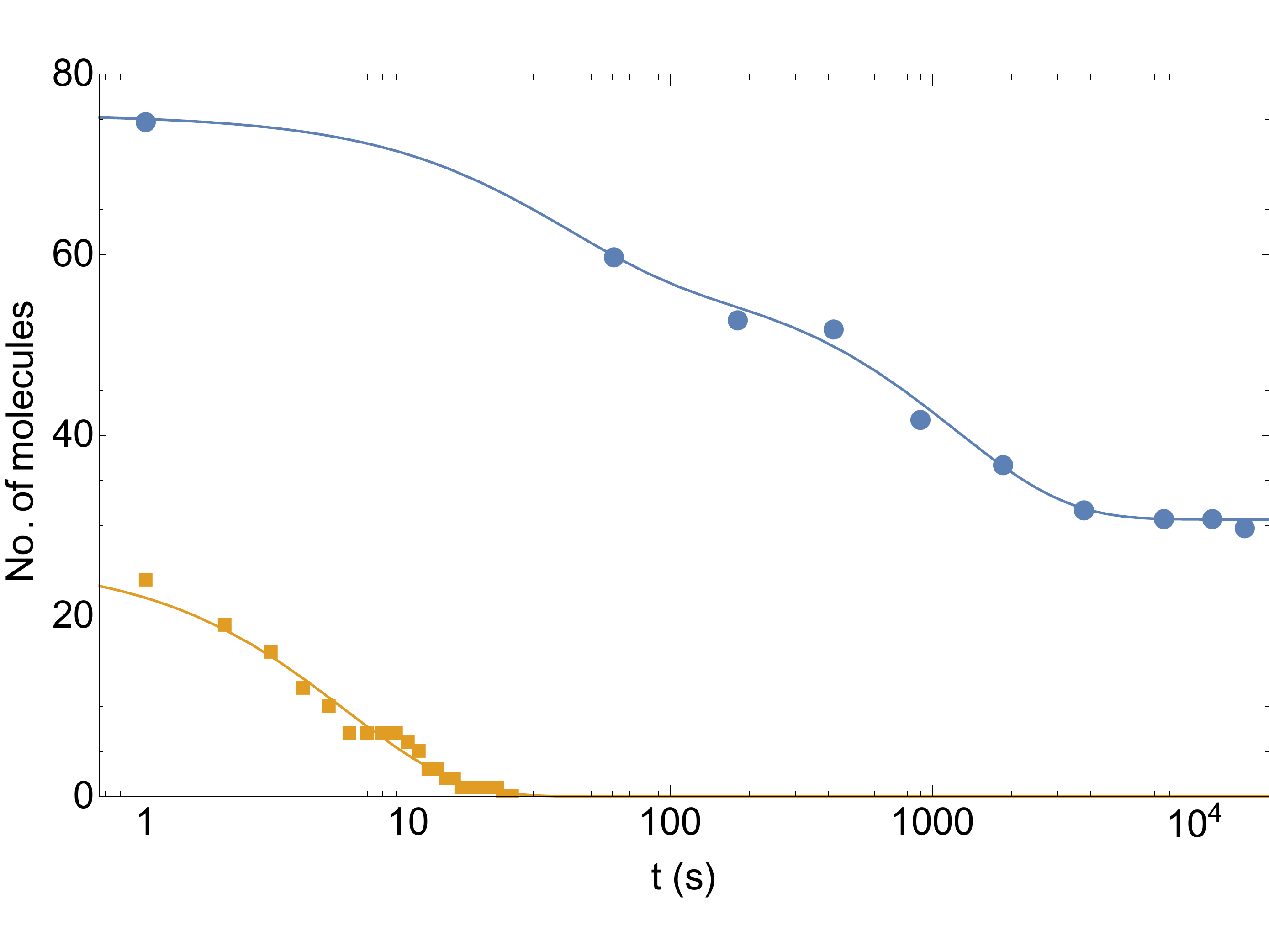}
\caption{Photo-bleaching of DBT in thin anthracene crystals. The graphs show the number of DBT molecules remaining unbleached as a function of the exposure time. Orange squares: Anthracene grown in an atmosphere of air and excited at 780 nm using an intensity of $10\,\mbox{kW/cm}^2$. Orange line: a fit to exponential decay, giving a lifetime of $5.7\,$s. Blue Circles: Anthracene grown in a nitrogen atmosphere and excited at 780 nm with $130\,\mbox{kW/cm}^2$. Blue line: simple empirical formula $15 e^{(-t/5.7)}+ 31 e^{(-t/10^3)}+ 30$ intended to show that the decay rate per molecule increases with time and hence that the bleaching is not exponential.}
\label{Fig:photobleach}
\end{figure}

When dye molecules are optically excited it is not unusual for them to react with ambient oxygen molecules and stop fluorescing \cite{Kozankiewicz:14}, which is clearly undesirable in a practical single-photon source. In order to see whether oxygen plays an important role here, we grew DBT-doped crystals in an atmosphere of air instead of purging the glove bag with nitrogen. We then illuminated 36 molecules, one at a time, at a wavelength of $780\,$nm and an intensity of $\sim10\,\mbox{kW/cm}^2$ ($s\simeq 0.13$) and recorded how long each molecule survived before photo-bleaching. The orange squares in Fig.~\ref{Fig:photobleach} show the number of molecules remaining unbleached as a function of the illumination time. This evolution is reasonably well described by an exponential decay, with a lifetime of $5.7\,$s, as shown by the solid line. That corresponds to a molecule being excited (on average) $\sim10^7$ times before it photo-bleaches. Nicolet \textit{et al.} \cite{Nicolet1:07} give the  branching ratio for decay from $S_1$ to the triplet state $T_0$  as $\sim10^{-7}$, so our result could indicate that the photobleaching mechanism here is associated with passing through the triplet state of DBT and subsequent reaction with the oxygen. However, DBT molecules on the surface of the slide, without any anthracene on top, bleach so quickly that they do not have time to pass through the triplet state -- indeed so quickly that we could not make a plot of their survival over time -- so, at least in that case, the triplet state of DBT does not seem to be playing a role in the bleaching.

By contrast, the DBT molecules prepared without oxygen are extremely photostable. In order to measure their bleaching, we removed LI and used L2 (shown in Fig.~\ref{Fig:oil_confocal}) to focus the excitation light onto the back aperture of the objective. This illuminated a circle of 9.3~$\mu$m FWHM in the plane of the sample, allowing us to expose all the molecules in a single crystal. We also used much higher laser power ($125~\mbox{mW}$ from a Toptica TA 100 Pro at $780\,$nm) so that the intensity on each molecule increased to $\sim 130$~kW/cm$^2$ ($s\simeq 1.7$). We then interleaved periods of high-intensity exposure with confocal scans to check how many molecules remained unbleached. The result, plotted using blue circles in Fig.~\ref{Fig:photobleach}, shows that these molecules are remarkably stable.  An initial population of 75 molecules, observed in 3 separate crystals, decays to 60 molecules after a minute, to 32 after an hour, and to 30 after 4 hours and 20 minutes. This evolution cannot be described by an exponential: the loss rate per molecule is much higher at early times than it is at later times. To illustrate this, the solid blue line in Fig.~\ref{Fig:photobleach} shows that this data can be described by the simple empirical formula $15 e^{(-t/5.7)}+ 31 e^{(-t/10^3)}+ 30$. At the end of this experiment, the 30 photostable molecules have survived $> 10^{12}$ excitations, and it seems likely that they could have survived many more. It remains to be seen whether all the molecules can be rendered photostable by a more serious effort to remove all the oxygen.

\section{Conclusions}
\label{sec:conclusions}

We have demonstrated a method for growing anthracene crystals that are wide and thin. The thickness, adjustable by the duration of the growth over the range $40 - 150\,$nm  (at least), has excellent uniformity. A highly supersaturated anthracene vapour is used to obtain enough chemical potential difference to favour this morphology. The anthracene is doped with DBT molecules that are spun onto the surface prior to the crystal growth, and this makes it possible to adjust the concentration of the dopant over a wide range. We have shown that after the anthracene is grown, the DBT molecules become aligned with a common axis in the crystal -- presumably the b-axis -- which indicates that they are accepted into the lattice, where most of them occupy the main insertion site. This opens the possibility of aligning the molecules with nano-fabricated optical structures when we seek in the future to insert them into devices. The standard Hanbury Brown and Twiss method has shown that the bright spots seen in the microscope are indeed single molecules, and the decay of the exponentials in the $C^{(2)}$ function have shown that the lifetime in these particular crystals is similar to that in other anthracene matrices. We have also measured the saturation intensity at room temperature and find that it is unremarkable. On subjecting these molecules to a long period of intense illumination, we find that they are extremely photostable as long as we grow the anthracene in a nitrogen atmosphere, rather than in air. We presume that this hinges on the level of oxygen.

In summary, there is a need for quantum emitters able to radiate individual photons on demand into optical nanostructures, and  we have started to pursue the proposal that DBT molecules in an organic matrix might be able to satisfy that need. Having shown that the matrix must be thin to achieve efficient coupling to a waveguide, we have found a way to grow suitable films of anthracene, and how to dope them with DBT. On investigating their optical properties, we find that these molecules are very promising, being both polarised and photostable. Whilst all of the anthracene growth presented in this paper was carried out on a glass substrate we have recently grown crystals on silicon nitride, a material commonly used for nanophotonic devices operating at visible wavelengths. We find that these crystals appear to be the same as those presented here.

\section{Acknowledgements}

We are indebted to Jaesuk Hwang for suggesting this method of crystal growth, to Kiang-Wei Kho for helping to set up the microscope, to Yu Hung Lien for valuable discussions, and to Jon Dyne, Steve Maine, and Valerijus Gerulis for their expert mechanical and electrical workshop support. We acknowledge funding from the Royal Society, the EPSRC, and dstl. The European Commission supports A.S.C. through a Marie Sk{\l}odowska Curie Individual Fellowship (Q-MoPS) and S.G. through the Action-Initial Training Network: Frontiers in Quantum Technology.

\end{document}